\documentclass[aps,prb,twocolumn,showpacs,superscriptaddress,amsmath]{revtex4}
\usepackage{graphicx}
\usepackage{amssymb}
\usepackage{citesort}

\begin{document}

\title{Photophysics of charge-transfer excitons in thin films of $\pi$-conjugated polymers}

\author{Zhendong Wang}
\affiliation{Department of Physics, University of Arizona
Tucson, AZ 85721}
\author{Sumit Mazumdar}
\affiliation{Department of Physics, University of Arizona
Tucson, AZ 85721}
\author{Alok Shukla}
\affiliation{Physics Department, Indian Institute of Technology, Powai, Mumbai 400076, India}
\date{\today}
\begin{abstract}
We develop a theory of the electronic structure and photophysics of interacting chains
of $\pi$-conjugated polymers to understand the differences between solutions and
films. 
While photoexcitation generates only the intrachain exciton in solutions, the optical 
exciton as well as weakly allowed charge-transfer excitons are generated in films. 
We extend
existing theories of the lowest polaron-pair and charge-transfer excitons to obtain descriptions of
the excited states of these interchain species, and show that a significant fraction
of ultrafast photoinduced absorptions
in films originate from the lowest charge-transfer exciton. Our proposed mechanism 
explains the simultaneous observation of polaronlike induced absorption features peculiar to
films in ultrafast spectroscopy and the absence of mobile charge carriers as deduced from
other experiments. We also show that there is
a 1:1 correspondence between the essential states that describe the photophysics of single chains
and of interacting chains that constitute thin films.
\end{abstract}
\maketitle

\section{Introduction}
The photophysics of dilute solutions and thin films of $\pi$-conjugated polymers (PCPs)
are often remarkably different 
\cite{Rothberg06,Arkhipov04,Schwartz03,Jenekhe94,Jenekhe95,Yan94,Hsu94,Martini04,Samuel95,Samuel98,Schweitzer99,Hertel01,Ho01,Lim02,Brown03,Koren03,Clark07,Sheng07,Singh08}.
It is generally
accepted that solutions exhibit
behavior characteristic of single strands, and the different behavior of films
is due to interchain interaction and disorder. Microscopic understanding
of the effects of interchain interaction has remained incomplete
even after intensive investigations. 
As discussed below, to a large extent this is because the experimental results
themselves, or their interpretations are controversial and confusing.
Theoretical investigations of the effects of interchain interactions 
\cite{Clark07,Mizes94,Wu97,Conwell98,Meng98,Ruini02,Cornil98,Beljonne00,Spano05,Barford07}
have until now focused largely on the lowest interchain species near the optical edge, 
and the role of such interchain species on the emissive behavior of films.
The goal of this work is to re-examine
interchain interaction in PCPs within a semi-empirical Hamiltonian with realistic
parameters, focusing on ultrafast photoinduced absorption (PA) measurements and 
related experimental results that appear to be mutually contradictory. We show that
theoretical understanding of {\it excited states} of interchain species is crucial for
this purpose.

The interchain species we will be interested in have been discussed by numerous
authors over the years, and the nomenclature has sometimes been confusing. It is therefore
important to fix the nomenclature before we begin. We will refer to intrachain 
neutral excitations
as {\it excitons}, independent of their binding energy. At the other extreme are the
{\it polaron-pairs}, which consist of two {\it completely ionic charged chains}, one positive
and one negative. Since we
consider nonzero interchain Coulomb interactions, and since we will be discussing 
two-chain systems only, the polaron-pair states are necessarily bound by Coulomb interactions.
For nonzero electron hopping
between the chains, eigenstates that are superpositions of the intrachain exciton 
and the polaron-pair are obtained. We will refer to these superpositions
as {\it charge-transfer
excitons}, hereafter CT excitons. The reader should note that the polaron-pair has been 
sometimes referred to as the
charge-transfer exciton in the literature. \cite{Wu97} 
The CT excitons, in their turn, have sometimes been
called excimers. \cite{Wu97,Conwell98,Meng98} 
Our nomenclature is based on the most
common useage of these terms, and we give precise quantum-mechanical definitions of these
interchain species in section IV.
One major difference between the work presented here and 
the existing literature is that we
are also interested in {\it higher energy excited polaron-pairs
and CT excitons}, which are defined exactly as above (thus, a high energy
CT exciton is predominantly a superposition of a similar high energy excited
intrachain exciton and polaron-pair). We find a 1:1 correspondence between
the ``essential states'' that determine the photophysics of
single strands \cite{Dixit91,Abe92,Lavrentiev99,Beljonne97,Shukla03} and the dominant excited states
including excited interchain species, that determine the photophysics of interacting chains.

Starting from a microscopic $\pi$-electron Hamiltonian,
we investigate the energy spectrum of interacting
PCP chains. 
We do not attempt to understand details of the photoluminescence (PL), which can be 
understood to a large extent within existing theories.
\cite{Clark07,Wu97,Conwell98,Meng98,Cornil98,Beljonne00,Spano05,Barford07}
Understanding delayed emission in PCP films (see below), on the other hand, will require 
much more sophisticated modeling.
We rather focus on the theory of excited state absorption in interacting chains,
with the goal of understanding the observed branching of photoexcitations and the origin
of the polaronlike photoinduced absorptions (PAs), \cite{Yan94,Hsu94,Sheng07,Singh08} and experiments that 
indicate that in spite of the occurrence of these polaronlike PAs free charges are
not generated as primary photoexcitations. \cite{Dicker04,Hendry05}

In the next section we present a brief yet detailed summary of relevant experiments in
PCP films that indicate the strong role of interchain interactions, highlighting
in particular the apparently contradictory observations. Following this, in
section III we present our theoretical model. In section IV we discuss the formation
of CT excitons and excited state absorptions from them, and present detailed
computational results. Finally, in section V we compare our theoretical results and
experiments, and present our conclusions. The computational results presented in section
III are for finite oligomers of PPV. In a separate Appendix we discuss the chain-length dependence 
of our results. We believe that our results apply to real materials.

\section{Review of Experimental Results}

PL from films is often redshifted relative to that from
dilute solutions, and the quantum efficiency (QE) of the PL from films 
is usually much smaller. PL from regioregular polythiophene (rrP3HT)
has recently been discussed within a weak-coupling H-aggregate model, within
which dipole-dipole coupling leads to an exciton band. \cite{Clark07} Absorption here
is to the highest state in the exciton band while emission is from the
lowest state.
\cite{Clark07,Cornil98,Spano05,Barford07} 
Conversely, it has
been claimed that PL from films of 
cyano-poly(paraphenylenevinylene), CN-PPV, and 
poly(2-methoxy,5-(2$^{\prime}$-ethyl-hexyloxy)1,4 paraphenylenevinylene), MEH-PPV, are from
CT excitons that occur below the intrachain optical exciton.
\cite{Samuel95,Samuel98,Wu97,Conwell98,Meng98} Formation of CT excitons requires that
polaron-pairs are energetically proximate to the excitons (see section IV and Appendix).
The occurrence of low energy polaron-pairs
is indicated by the observation of ``persistent'' or delayed PL lasting until
milliseconds in films, the electric field quenching of the delayed PL, and the resumption
of the PL upon removal of the field. 
\cite{Rothberg06,Arkhipov04,Schweitzer99,Hertel01} 

Experiments that also indicate the strong role of interchain
interactions, and that are even more difficult to understand than PL involve transient
absorption. Two distinct ultrafast photoinduced absorptions (PAs)  
are seen in solutions as well as in films with weak
interchain interactions, such as dioctyloxy-poly-paraphenylenevinylene 
(DOO-PPV). \cite{Frolov00,Frolov02}
The low energy PA$_1$ appears at a threshold energy of 0.7 eV and has a peak at
$\sim$ 1 eV, while the higher energy PA$_2$ occurs at $\sim$ 1.3--1.4 eV. Comparison
of PA and PL decays \cite{Frolov00,Frolov02} and other nonlinear spectroscopic measurements 
\cite{Liess97} have confirmed that these PAs are from the 1B$_u$
optical exciton, in agreement with theoretical work on 
PCP single chains. \cite{Dixit91,Abe92,Lavrentiev99,Beljonne97,Shukla03}
In contrast, PA and PL in PCPs with significant
interchain interactions (for e.g., 
MEH-PPV) are 
uncorrelated. \cite{Rothberg06,Yan94,Hsu94} It has been argued that PAs in such systems
is from the polaron-pair. \cite{Rothberg06,Yan94,Hsu94} This would require
generation of the
polaron-pair in ultrafast time scales. The
mechanism by which such ultrafast generation can occur is not clear. \cite{Basko02} The
possibility that the PAs here are from the CT exciton has not been theoretically investigated.

Recent experiments have contributed further to the mystery.
Sheng {\it et al.} have extended femtosceond (fs) PA measurements
to previously inaccessible wavelengths, \cite{Sheng07} and
have detected two additional weak PAs in film samples of MEH-PPV, PPV and rrP3HT that are
absent in solutions of the PPV derivatives as well as in regiorandom (rraP3HT),
which is known to have weaker interchain interaction than rrP3HT. The authors initially
assigned the new low energy PA at $\sim$ 0.35--0.4 eV, labeled P$_1$, and the higher
energy PA, labeled P$_2$ in this work, to absorptions of free polarons that according
to the authors are generated when interchain interactions are strong. 
The high energy PA associated with films had been previously observed in
MEH-PPV,
\cite{Rothberg06,Yan94,Hsu94} and it has been ascribed to absorptions from free polarons
as well as from polaron-pairs. \cite{Mizes94}
Interestingly,
these PAs peculiar to films are generated instantaneously, suggesting branching of 
photoexcitations with competing channels generating excitons and polarons.
Such branching of photoexcitations would be in agreement
with previous claim of the observation of infrared active vibrations (IRAV) in MEH-PPV
in fs time, \cite{Miranda01} but is difficult to reconcile with
the large exciton binding energies deduced from PA$_1$ energy, 
\cite{Frolov00,Liess97,Dixit91} which requires that polarons are generated from
dissociation of the exciton due to extraneous influence at a later time. 
Instantaneous IRAV
\cite{Miranda01} has not been observed by other experimentalists, and interpretations other
than those given by the original authors exist in the literature. \cite{Stevens01}
In the context of Sheng {\it et al's} experiment, the following is, however, true: 
if the P$_1$ absorption, as well as the high
energy absorption absent in solutions are indeed due to polarons, IRAV associated with these
absorptions should have been observed.
Intriguingly,
Sheng {\it et al.} in their experiments did not find any IRAV at room temperatures
that should have accompanied the P$_1$ absorption, and very weak IRAV at
80 K.
\cite{Sheng07} Later more careful attempts have also failed to detect room temperature
IRAV. \cite{Vardeny08}

The absence of room temperature IRAV suggests that polarons are {\it not} being generated 
in Sheng {\it et al's} experiment. This conclusion is in apparent agreement with
microwave conductivity measurements
\cite{Dicker04} and THz spectrscopy \cite{Hendry05} that have
found negligible polaron generation upon direct photoexcitation
in both solutions and films.
Based on very recent experiment that probed the polarization
memory decay of photoexcitations, Singh {\it et al.} have concluded that the high
energy PA associated with films is not from free polarons but 
from a CT exciton
(note: these authors use the terminologies CT exciton and excimer
synonymously).
\cite{Singh08} 
The origin of the polaronlike features in the PA thus remains mysterious.

\section{Theoretical model}

Our calculations are within an extended two-chain Pariser-Parr-Pople Hamiltonian 
\cite{Pariser53,Pople53}
$H = H_{intra}+H_{inter}$, where $H_{intra}$ and $H_{inter}$ correspond to intra-
and interchain components, respectively. $H_{intra}$ is written as,
\begin{multline}
\label{H_PPP}
H_{intra} = - \sum_{\mu, \langle ij \rangle, \sigma} t_{ij}
(c_{\mu,i,\sigma}^\dagger c_{\mu,j,\sigma}+ H.c.) + \\
U \sum_{\mu,i} n_{\mu,i,\uparrow} n_{\mu,i,\downarrow} 
+ \sum_{\mu,i<j} V_{ij} (n_{\mu,i}-1)(n_{\mu,j}-1)
\end{multline}

where $c^{\dagger}_{\nu,i,\sigma}$ creates a $\pi$-electron of spin $\sigma$ on
carbon atom $i$ of oligomer $\nu$ ($\nu$ = 1, 2), $n_{\nu,i,\sigma} = 
c^{\dagger}_{\nu,i,\sigma}c_{\nu,i,\sigma}$ is
the number of electrons on atom $i$ of oligomer $\nu$ with spin $\sigma$ and
$n_{\nu,i} = \sum_{\sigma}n_{\nu,i,\sigma}$ is the total number of electrons on atom
$i$. The hopping matrix element
$t_{ij}$
is restricted to nearest neighbors and
in principle can contain electron-phonon interactions, although
a rigid bond approximation is used here. $U$ and $V_{ij}$ are the
on-site and intrachain intersite Coulomb interactions.
We parametrize $V_{ij}$
as \cite{Chandross97}
\begin{equation}
\label{parameters}
V_{ij}=\frac{U}{\kappa\sqrt{1+0.6117 R_{ij}^2}} 
\end{equation}
where $R_{ij}$ is the distance between carbon atoms $i$ and $j$ in
\AA, and $\kappa$ is the dielectric
screening along the chain due to the medium. Based on previous work \cite{Chandross97}
we choose $U$ = 8 eV and $\kappa$ = 2.
We write $H_{inter}$ as
\begin{align}
\label{inter}
H_{inter} &= H_{inter}^{1e}+H_{inter}^{ee} \\
H_{inter}^{1e} &= -t_{\perp}\sum_{\nu <
\nu^{\prime},i,\sigma}(c^{\dagger}_{\nu,i, \sigma}
c_{\nu^{\prime},i,\sigma} + H.C.)  \\
H_{inter}^{ee}&=\frac{1}{2}\sum_{\nu < \nu^{\prime},i,j} V_{ij}^{\perp}(n_{\nu,i} -1)(n_{\nu^{\prime},j} - 1)
\end{align}

We will assume planar cofacial stacking of oligomers in our calculations. While such ideal
stacking does not occur in real systems, it is believed that this assumption captures the
essential physics of polymer films. \cite{Clark07,Wu97,Conwell98,Meng98,Cornil98,Beljonne00,Spano05,Barford07}
In the above, $t_{\perp}$ is restricted to
nearest interchain neighbors.
We choose $V_{ij}^{\perp}$ as in Eq. \ref{parameters}, with
a background dielectric constant $\kappa_{\perp} \leq \kappa$. \cite{Conwell98}

\section{Charge-transfer excitons}

\subsection{Coupled ethylenes}

In order to get a physical understanding
of the effect of $H_{inter}$, we begin with the case of two ethylene molecules,
placed cofacially one on top of the other such that the overall structure has a center of inversion.
The small number
of energy states here permit clear identification of all two-chain excitations. Although
full configuration interaction (FCI) can be be performed in this case, in view of our 
interest in long PPV oligomers, we will restrict our calculations as well as physical
discussions to the single configuration interaction (SCI) approximation
(see, however, section IV.D.) 

We consider first the $U=V_{ij}=0$ molecular orbital (MO) limit for $H_{intra}$.
The ethylene MOs are written as,
\begin{equation}
\label{a_dagger}
a_{\nu,\lambda,\sigma}^\dagger = {1 \over \sqrt{2}}
[c_{\nu,1,\sigma}^\dagger + (-1)^{(\lambda - 1)}c_{\nu,2,\sigma}^\dagger]
\end{equation}
where $\lambda=1(2)$ corresponds to the bonding (antibonding) MO.
The spin singlet one-excitation space for the two molecules consists of 
four configurations. Two of these four configurations consist of neutral molecules, 
with either of the two molecules excited and the other in the ground state; the other 
two consist of positively and negatively charged molecules, with each charged molecule
in its lowest state.
We refer to the neutral configurations with
intramolecular excitations as excitons, and write them as
$|exc1 \rangle$ and $|exc2 \rangle$,
ignoring for the moment that 
true excitons require nonzero $U$ and $V_{ij}$. We will refer to the charged molecule
pair as polaron-pairs and write them as
$|P_1^+P_2^- \rangle$ and $|P_1^-P_2^+ \rangle$, respectively.
The exciton and polaron-pair wavefunctions are given by,
\begin{align}
\label{VB}
|exc1 \rangle = {1 \over \sqrt{2}}a_{2,1,\uparrow}^\dagger a_{2,1,\downarrow}^\dagger (a_{1,1,\uparrow}^\dagger a_{1,2,\downarrow}^\dagger -
a_{1,1,\downarrow}^\dagger a_{1,2,\uparrow}^\dagger)|0 \rangle \\
|P_1^+P_2^- \rangle = {1 \over \sqrt{2}}a_{2,1,\uparrow}^\dagger a_{2,1,\downarrow}^\dagger (a_{1,1,\uparrow}^\dagger a_{2,2,\downarrow}^\dagger -
a_{1,1,\downarrow}^\dagger a_{2,2,\uparrow}^\dagger)|0 \rangle 
\end{align}
The terms within the parenthesis in Eqs (7) and (8) constitute singlet bonds between MOs.
The basis functions $|exc2 \rangle$ and $|P_1^-P_2^+  \rangle$ are obtained from the above
by applying mirror-plane symmetry.  
The four spin-bonded valence bond (VB) diagrams corresponding to the excitons and
polaron-pairs are shown in Fig.~1(a).

Nonzero $H_{inter}$ mixes these pure states
to give the CT excitons, \cite{Wu97,Conwell98,Meng98} the theory of which is fundamentally 
similar to the
Mulliken's theory of ground state charge-transfer \cite{Mulliken52} 
(except that the excited state 
Hamiltonian for identical molecules involves four instead of two basis functions).
Consider first the $H_{inter}^{1e}=0$ limit.
Matrix elements of $H_{inter}^{ee}$ are zero between $|P_1^+P_2^- \rangle$ and 
$|P_1^-P_2^+ \rangle$ but nonzero between $|exc1 \rangle$ and $|exc2 \rangle$, 
indicating that
while the polaron-pair states are degenerate for $H_{inter}^{ee} \neq 0$,
the exciton states form new nondegenerate states $|exc1 \rangle \pm |exc2 \rangle$
(see Fig.~1(b)).
The dipole operator $\pmb{\mu} = e\sum_{\nu,i}\pmb{r}_{\nu,i}(n_{\nu,i}-1) $, 
where $\pmb{r}_{\nu,i}$ gives the location of atom
$i$ on oligomer $\nu$,
couples the ground state to only the even parity 
exciton state. The odd parity exciton is now a dark state occurring below the
optical exciton. The splitting of the exciton states due to Coulomb interactions alone
can be described within the dipole-dipole approximation. \cite{Clark07,Cornil98,Beljonne00,Spano05,Barford07}

We now switch on $H_{inter}^{1e}$,
which mixes the odd parity neutral $|exc1 \rangle-|exc2 \rangle$ and charged 
$|P_1^+P_2^- \rangle-|P_1^-P_2^+ \rangle$,
to give the two CT exciton states in Fig.~1(c). 
The extent of configuration mixing 
depends on the relative energy separation between the pure polaron-pair and the odd parity 
exciton in Fig.~1(b) and the
magnitude of $H_{inter}^{1e}$, {\it i.e.}, on $V_{ij}^{\perp}/t_{\perp}$. 
For significant $V_{ij}^{\perp}$ (attraction between interchain electron and hole), the
polaron-pair can be low in energy (see Appendix).
The CT excitons, being superpositions of the dark exciton state and
odd parity polaron-pair configurations, neither of which are accessible in intramolecular 
optical excitation, are optically forbidden from the ground state.
The even parity states, the  optical exciton $|exc1 \rangle + |exc2 \rangle$
and the polaron-pair $|P_1^+P_2^- \rangle+|P_1^-P_2^+ \rangle$, are not affected
by $H_{inter}^{1e}$ in this symmetric case.

\begin{figure}
 \centering
 \includegraphics[clip,width=3.375in]{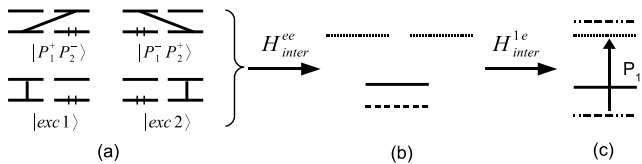}
 \caption{(a). The one-excitation space of two weakly
interacting oligomers. For each oligomer one bonding and one antibonding MO is shown.
The MOs can be occupied by 0, 1 or 2 electrons. The singly occupied MOs are connected 
by singlet bonds. (b) and (c). Eigenstates of
$H_{inter}^{ee}$ and total $H_{inter}$, respectively.
Solid lines, even parity exciton; dashed line, odd parity
exciton; dot-dashed lines, CT excitons ; dotted lines, polaron-pairs.
The P$_1$ induced absorption is indicated in (c).
}
\end{figure}

We now make an observation that will be central to the work presented in the next sections: 
matix elements of the {\it transverse} component 
of $\pmb{\mu}$, perpendicular to the molecular axes, between the CT excitons and the
even-parity polaron state are nonzero, and proportional to $t_{\perp}$. For nonzero
$t_{\perp}$ we therefore expect {\it excited state charge-transfer absorption} 
from the CT exciton
to the polaron-pair state, the strength of which is
proportional to $t_{\perp}^2/\Delta E$, where $\Delta E$ is the energy difference between the
initial and final states. \cite{Mulliken52} This is shown explicitly in the next section.
We label this photoinduced CT absorption
as P$_1$, as we will show that it is this induced absorption in the context of PCPs that
corresponds to the 
PA labeled P$_1$ by Sheng {\it et al.}. \cite{Sheng07}

The above discussion starts from the $U=V_{ij}=0$ limit of $H_{intra}$
only for simplicity. For the particular case of two ethylenes, the two neutral
exciton configurations are the same, independent of $U$ and $V_{ij}$. 
The description of the polaron-pair states also remains the same within the SCI approximation
(higher energy two electron-two hole excitations can modify the polaron-pair states in
approximations that go beyond SCI). 
For nonzero $U$ and $V_{ij}$, the dominant contribution to the stabilization of the 
lower CT exciton still comes from the
configuration mixing with the odd-parity polaron-pair basis function. The
strength of the dipole-coupling between the CT exciton and the even-parity pure
polaron-pair state, the P$_1$ absorption in Fig.~1(c), is again $t_{\perp}^2/\Delta E$, 
where, however, $\Delta E$ now depends on $U$ and $V_{ij}$.

\subsection{Cofacial PPV oligomers, symmetric case}

We now go beyond the two ethylenes and make the following observations.

(i) Eqs.~7 and 8, or equivalently, the four spin-bonded VB diagrams in Fig.~1(a), with
the MOs corresponding to the highest occupied and lowest unoccupied MOs (HOMOs and LUMOs), 
describe the lowest intramolecular excitations and the lowest energy
polaron-pair states for arbitrary PCP oligomers in the $U=V_{ij}=0$ limit of $H_{intra}$. 
For extension of the above concept to arbitrary PCPs,
only Eq.~6 needs to be modified, with the MOs now superpositions of a larger number
of atomic wavefunctions. The degenerate neutral exciton states are once again split
by the Coulomb interactions in $H_{inter}$ alone, even for $t_{\perp}=0$, as in Fig.~1(b). 
This is the
basis for the dipole-dipole approximation to exciton splitting.  
\cite{Clark07,Cornil98,Beljonne00,Spano05,Barford07}

The
dipole-dipole approximation, or more precisely, the $t_{\perp}=0$ approximation, 
ignores the CT between the odd parity polaron-pair and exciton states. The extent of
this CT, as pointed out in the above, depends
on the magnitude of the {\it effective} electron hopping between the MOs of the two interacting
oligomers, and the energy separation between the pure odd parity polaron-pair and
exciton states in the $t_{\perp}=0$ limit. 
The energy separation between
the polaron-pair and the exciton depends on the difference in the
electron-hole separations in the intrachain exciton and the polaron-pair, a quantity difficult to
evaluate from first principles. 
The relative energy difference between the polaron-pair and the
exciton in theoretical work can therefore be based only on interpretations
of experiments. The likelihood of short
electron-hole separations in the polaron-pair (and therefore energy close to the
exciton, which in turn increases CT) has been suggested by several authors. 
Based on the work reported in references \onlinecite{Rothberg06} and
\onlinecite{Arkhipov04}, Wu and Conwell, \cite{Wu97,Conwell98} and
Meng \cite{Meng98} have previously assumed low energy polaron-pair states, and have
described the CT process in PPV derivatives
within a simplified $H^{ee}_{inter}$, in order to explain the reduced PL in films
(indeed, references \onlinecite{Schweitzer99} and \onlinecite{Hertel01} suggest that
a fraction of the polaron-pair states occur even {\it below} the exciton).
We have found in our calculations that for $k_{\perp} \leq 2.5$ (see Eqs. 2 and 5)
the fundamental
assumption of Wu and Conwell and Meng continues to be valid for long chains of 
polyenes and PPVs (see Appendix), and we adopt the same approach. 

To conclude, Fig.~1
applies to the $U=V_{ij}=0$ limit of arbitrary PCPs, when the excitations involve the
HOMO and LUMO of the two identical cofacial oligomers. We have verified this from
CI calculations with $H_{intra}=0$ but $H_{inter} \neq 0$ for long PPV oligomers. 
The energy splitting between the exciton and
the CT exciton is relatively insensitive to chain length.
To identify wavefunctions as polaron-pair, CT exciton, etc., 
we choose an  orbital set consisting of the 
Hartree-Fock orbitals of the individual molecular
units, and perform CI calculations using these localized MOs.
The localized basis allows calculations of ionicities of individual oligomers.
The expected ionicities are 0 and 1 for the exciton and the polaron-pair, respectively,
and fractional for the CT excitons. 

(ii) There is no {\it a priori} reason to assume that the MOs in Fig.~1 should include
only the HOMOs and the LUMOs of the PCP oligomers. {\it Higher energy excited
exciton and polaron-pair states}, involving bonding (antibonding) MOs 
below (above) the HOMO (LUMO), can also be coupled by $H_{inter}$, provided once again,
the excited polaron-pairs and the excitons are close in energy. 
Again, we have confirmed this from CI calculations in the $H_{intra}=0$, $H_{inter} \neq 0$
limit for PPV oligomers, using the localized basis.

\begin{table}
\caption{SCI excited states of two symmetrically
placed 8-unit PPV oligomers for $\kappa_{\perp}=2$,
$t_{\perp}=$ 0.1 eV.
Here $j$ and $E_j$ are quantum numbers (without considering symmetry)
and energy, respectively. Ionicity is the charge on the
chains. 
The states are arranged not according to their energies, but 
according to the manifolds they belong to (see text). The
$\mu_{G,j}$ and $\mu_{i,j}$ are the dipole couplings (electronic
charge = 1)
between the ground state and state $j$, and between excited states,
respectively.}
\begin{ruledtabular}
\begin{tabular}{ccccc}
   $j$ & $E_j$ (eV) & Ionicity  & $\mu_{G,j}$ & $\mu_{i,j}$ \\ \hline
  2	   & 2.67 & 0.26 & 0    & \textemdash \\
  4    &  2.81 &     0    & 6.52 & \textemdash       \\
  5    & 3.00 &   1  & 0 & 2.04$^a$     \\
  8    & 3.12 & 0.74 &  0   & \textemdash       \\ 
  \\
  3	   & 2.81 & 0.29 & 0  & \textemdash \\
  7    & 3.06 & 0       & 0  & \textemdash   \\
  9    & 3.12  & 1   &  0 & \textemdash  \\
 10   & 3.24 & 0.64& 0  & \textemdash      \\ 
  \\
 11   & 3.26 & 0.38 & 0  & 6.91$^b$\\
 15   & 3.42 & 0     & 0  & 6.83$^{b,c}$\\
 19   & 3.46 & 1   &  0 & 7.68$^b$\\
26   & 3.67 & 0.55 & 0  & 6.69$^b$\\  
\end{tabular}
\end{ruledtabular}
\footnotetext{$^a$i=2. $^b$All dipole couplings are with states in lowest
manifold near 1B$_u$ with the same character (see text). $^c$ The mA$_g$.}
\end{table}

\begin{table}
\caption{SCI eigenstates of cofacial PPV oligomers of lengths 7 and 9-units,
respectively, with only one end matching. All parameters are the same as in
Table I. The classifications of states in the last column are obtained from
wavefunction analysis (see text)}
\begin{ruledtabular}
\begin{tabular}{cccccc}
   $j$ & $E_j$ (eV) & Ionicity  & $\mu_{G,j}$ & $\mu_{i,j}$ & state type \\ \hline
  2     & 2.67 & 0.25 & 0.78    & \textemdash & CT exciton \\
  3     & 2.81 & 0.11 & 4.99 & \textemdash & exciton      \\
  4     & 2.87 & 0.16 & 4.17 & \textemdash & exciton      \\
  6    & 3.01  & 1.00 & 0.00 &  2.03$^a$   &  polaron-pair  \\
  7     & 3.13 & 0.31 &  0.19   & \textemdash  & CT exciton   \\
  \\
  5     & 2.99 & 0.13 & 0.04  & \textemdash & two-photon exciton\\
  8     & 3.13 & 0.68 & 0.09  & \textemdash & CT exciton \\
  9     & 3.15 & 0.99  & 0.00 & \textemdash   & polaron-pair \\
  10    & 3.20 & 0.15  & 1.42  & \textemdash  & CT exciton    \\
  \\
  11    & 3.28 & 0.38 & 0.00  & 6.75$^b$ &   mA$_g$ CT exciton \\
  15    & 3.41 & 0.16 & 0.00  & 6.30$^{b}$ &  mA$_g$ exciton \\
  17    & 3.47 & 0.19 & 0.00  & 5.96$^{b}$  &  mA$_g$ exciton \\
  18    & 3.50 & 0.69  & 0.24 & 4.77$^b$   &  mA$_g$ ``polaron-pair'' \\
  19   & 3.52 & 0.62 & 0.76  &  3.42$^b$   &  mA$_g$ ``polaron-pair'' \\
\end{tabular}
\end{ruledtabular}
\footnotetext {$^a$i=2. $^b$All dipole couplings are with states in lowest
manifold near 1B$_u$ with the same character. }
\end{table}

(iii) The results of (i) and (ii) indicate that for $U=V_{ij}=0$ but
$H_{inter} \neq 0$, the two-chain energy spectrum consists of a series of 
overlapping energy manifolds, with each manifold containing an
exciton, a polaron-pair and two CT excitons, as in Fig.~1(c). 
For nonzero $U$ and $V_{ij}$, single chain excited eigenstates are
superpositions of the single chain MO configurations.
It is therefore reasonable to speculate that the two-chain spectrum for
{\it nonzero} $U$ and $V_{ij}$ also consists of similar energy manifolds, at
least upto the continuum band.
We have verified this, using the localized MO basis set and the SCI approximation,
including {\it all} one-excitations,
within the complete two-chain Hamiltonian.
We have summarized our results for two interacting
symmetrically placed cofacial 8-unit PPV oligomers at a distance of 0.4 nm in Table I, where
we have clearly indicated the different energy manifolds.
Intrachain one- or two-photon excitons, interchain polaron-pairs and CT excitons within each
manifold are easily identified from their ionicities and transition dipole couplings,
even at higher energies. 

\subsection{Cofacial PPV oligomers: unsymmetric case}

We now relax the inversion symmetry condition to take 
disorder into account approximately. This is important, as with nonzero interchain
hopping, it is not obvious that the characterizations of eigenstates as
intrachain excitons, CT excitons and polaron-pairs continue to be true at higher energies
in the absence of perfect symmetry. Furthermore,
we will see that such disorder also accounts for the appearance of instantaneous
signature of the PA P$_1$. \cite{Sheng07}
We consider cofacial oligomers of different lengths, 
with only one end matching (see insert, Fig.~2).
In Table II we show the
results of SCI calculations for PPV oligomers 7- and 9-units long, 
0.4 nm apart. We have verified that these results are independent of the actual lengths
of the oligomers, by performing similar calculations for pairs of oligomers of different
lengths ranging from 5 to 10 units. Unlike Table I, here we have given also the dominant character,
intrachain exciton, CT exciton or polaron-pair of each eigenstate.

As indicated in Table II, in the absence of inversion symmetry, characterizations
of eigenstates requires going beyond ionicities. 
We determine the dominant characters the eigenstates 
from detailed wavefunction analysis. For example, the $j=2$ state in Table II has large
overlaps with the odd parity configurations ${|exc1\rangle - |exc2\rangle}$ and 
${|P^+_1P^-_2\rangle - |P^+_2P^-_1\rangle}$, and very weak overlap with the
even parity ${|exc1\rangle + |exc2\rangle}$, identifying this state as predominantly
a CT exciton. Exactly the opposite is true for states $j$ = 3 and 4, which 
are the intrachain excitons.
The states $j=3$ and 4 also have weak but nonzero overlaps with the even parity 
${|P^+_1P^-_2\rangle + |P^+_2P^-_1\rangle}$, which gives them weak ionic character.
The occurrence of two distinct optical exciton states,
split by a very small energy difference, is a consequence of asymmetry. This characterization 
is in agreement with their large dipole couplings to the ground state, as well as small
ionicities. Another consequence
of asymmetry is that the CT exciton is now weakly dipole-coupled to the ground state
(the relative contributions to the wavefunction of $|exc1\rangle$  and $|exc1\rangle$ 
are unequal),
indicating weak but direct photogeneration of this state from the ground state.
The polaron-pair state in the lowest manifold ($j=6$) can be still identified by its ionicity
alone. Wavefunction analysis here indicates this state to be an even superposition
of $|P^+_1P^-_2\rangle$  and $|P^-_1P^+_2\rangle$. 
Exactly as in Table I
we find nonzero transition dipole-coupling between the $j=1$ CT exciton and the polaron-pair,
with the magnitude of the coupling nearly the same. Furthermore, the CT exciton continues
to have zero transition dipole coupling with all other states in this manifold.

The characterizations of the states in the second and third energy manifolds are obtained
similarly from calculations of overlaps with the fundamental basis functions. These
basis functions, however, involve higher energy single-particle excitations orthogonal
to those contributing to the states in the lower manifold (for example, the excitonic
basis functions contributing to the $j=5$ state in Table II has strong contributions
from the HOMO $\to$ LUMO+2 and HOMO -- 1 $\to$ LUMO contributions of each chain, identifying
it as a two-photon exciton.) The energy orderings within the manifolds can also
be different from that in the lowest manifold. Thus 
the ordering of the lowest intrachain and CT excitons are reversed in the
second manifold, with the lower energy $j=5$ being the intrachain exciton and the higher
energy $j=8$ being the CT exciton. We comment on the states labeled mA$_g$ in the
third manifold in Table II in the next subsection. Here we only point out that the
$j=18$ and 19 states, in spite of their intermediate ionicities, are predominantly
polaron-pair, based on their strong
overlaps with even superpositions of high energy charged configurations. 
The strong mixing between intrachain and interchain
basis functions in this region is a signature that this energy region is close to the
continuum band. \cite{Dixit91}

To summarize this subsection, characterizations of eigenstates as predominantly intrachain
exciton, CT exciton and polaron-pair continues to be valid even in the presence of disorder,
although they become less appropriate at higher energies. 

\subsection {Photoinduced absorptions}

Besides energies and ionicities, Tables I and II also list the transition dipole couplings
of excited states with the ground state, and between the excited states themselves.
The key results 
of Tables I and II are: (i) direct photogenerations of the
optical exciton and the two lowest CT excitons, one below and one above the intrachain exciton,
are allowed in the presence of disorder, and (ii) the lowest CT exciton
plays a crucial role in PCP films. 
We have verified that our results remain
qualitatively intact for three or more oligomers, different relative orientations and 
distances. 

The intrachain exciton states in the third manifolds, $j=15$ in Table I and $j=15$ and 17
in Table II correspond to the single chain mA$_g$ exciton, which is the two-photon
state that dominates single-chain photophysics.\cite{Dixit91,Abe92,Lavrentiev99,Beljonne97}
PA$_1$ in solutions is to the
mA$_g$. \cite{Shukla03} Our calculations indicate that exactly
as the transition dipole coupling is large between the single-chain 1B$_u$ and
the single-chain mA$_g$, \cite{Dixit91} equally large dipole couplings occur between pairs of states
in the 1B$_u$ and mA$_g$ manifolds that are of the same character (for example, from the
CT exciton in the 1B$_u$ manifold to the CT exciton in the mA$_g$ manifold). 

In Fig.~2 we compare PAs calculated for a single
8-unit PPV oligomer with that from the lowest CT exciton in a two-chain system
consisting of a 7-unit and a 9-unit oligomer. We have shown results for three different parameter sets
to indicate the relative insensitivity of our results to parameters. We performed similar calculations 
for many other combinations of chain lengths involving PPV oligomers of lengths from 5 to 10 units.
There is very little difference between the different cases (except that in the symmetric cases
the CT exciton has zero transition dipole coupling with the ground state).
PA$_1$ in the single chain corresponds 
to the transition from the 1B$_u$ to the
mA$_g$. The initial and final states of PA$_1^{\prime}$ absorptions in the two-chain
systems are both CT excitons. The P$_1$ absorption, missing in the single chain,
is from the lowest CT exciton to the lowest polaron-pair. The calculated PAs for the
two-chain system are {\it excited state equivalents} of the absorptions expected
within the classic Mulliken theory of weak donor-acceptor
complexes. In a donor-acceptor complex, there occurs weak CT absorption at low energy,
in addition to the molecular absorptions. \cite{Mulliken52}
In Fig.~2, P$_1$ is the CT absorption and
PA$_1^{\prime}$ the molecular absorption.

\begin{figure}
 \centering
 \includegraphics[clip,width=2.4in]{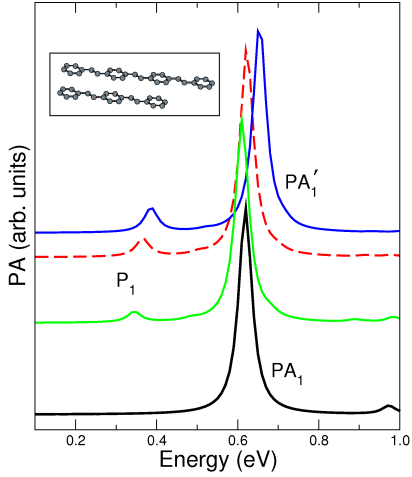}
 \caption{(Color online) Calculated PAs for a 8-unit PPV oligomer (black curve),
and for a two-chain system consisting of a
7-unit and a 9-unit oligomer: for $\kappa_{\perp}=2$ and $t_{\perp}=0.1$ eV (green);
$\kappa_{\perp}= 1.75$ and $t_{\perp}=0.15$ eV (red); and
$\kappa_{\perp}= 1.5$ and $t_{\perp}=0.2$ eV (blue).
A linewidth of 0.02 eV is assumed in all cases.
The inset shows schematically the arrangement of
the oligomers,
with the ends matching on one side only.
PA$_1$ is from the exciton. P$_1$ and PA$_1^{\prime}$ are from the CT exciton.
}
\end{figure}

Sheng {\it et al.}, \cite{Sheng07} and more recently, Singh {\it et al.} \cite{Singh08} have
also discussed a higher energy PA, above PA$_1^{\prime}$, peculiar to films. It is believed
\cite{Singh08} that this high energy PA is the same that was observed very early
in MEH-PPV. \cite{Yan94,Hsu94} As discussed
before in the context of the PA$_2$ absorption in single chains, \cite{Shukla03} such high 
energy regions cannot be investigated within the SCI approximation, and higher
order CI calculations become essential. Such calculations for pairs of PPV oligomers is
beyond our capability currently. On the other hand, as emphasized in section IV.A,
the photophysics of the
CT exciton in PCPs can be anticipated even from the behavior of coupled ehtylene molecules.
We have performed FCI calculations for pairs of ethylene and butadiene molecules and
have indeed detected a high energy two-electron two-hole polaron-pair state 
$|P^{+}_1P^{-}_2 - P^{+}_2P^{-}_1\rangle_{2e-2h}$ to which absorption from the
CT exciton is allowed. In the notation of section III, the components of this state 
for  the coupled ethylene system are,
\begin{multline}
\label{VB2}
|P_1^+P_2^- \rangle_{2e-2h}={1 \over 2}[a_{2,2,\uparrow}^\dagger a_{2,2,\downarrow}^\dagger 
(a_{1,1,\uparrow}^\dagger a_{2,1,\downarrow}^\dagger -
a_{1,1,\downarrow}^\dagger a_{2,1,\uparrow}^\dagger) \\
+ a_{2,1,\uparrow}^\dagger a_{2,1,\downarrow}^\dagger 
(a_{1,2,\uparrow}^\dagger a_{2,2,\downarrow}^\dagger -
a_{1,2,\downarrow}^\dagger a_{2,2,\uparrow}^\dagger)]|0 \rangle
\end{multline}
\begin{multline}
\label{VB3}
|P_2^+P_1^- \rangle_{2e-2h}={1 \over 2}[a_{1,2,\uparrow}^\dagger a_{1,2,\downarrow}^\dagger 
(a_{1,1,\uparrow}^\dagger a_{2,1,\downarrow}^\dagger -
a_{1,1,\downarrow}^\dagger a_{2,1,\uparrow}^\dagger) \\
+ a_{1,1,\uparrow}^\dagger a_{1,1,\downarrow}^\dagger 
(a_{1,2,\uparrow}^\dagger a_{2,2,\downarrow}^\dagger -
a_{1,2,\downarrow}^\dagger a_{2,2,\uparrow}^\dagger)]|0 \rangle
\end{multline}
Each polaron-pair configuration now has two components, related by electron-hole symmetry.
In Figs.~3(a) and (b) we have shown the four spin bonded VB diagrams that describe
this high energy polaron-pair state. The diagrams are similar for coupled butadienes,
with the only difference that there occur now bonding (antibonding) MOs below (above) the 
bonding (antibonding) MOs of Fig.3.

\begin{figure}
 \centering
 \includegraphics[clip,width=2.4in]{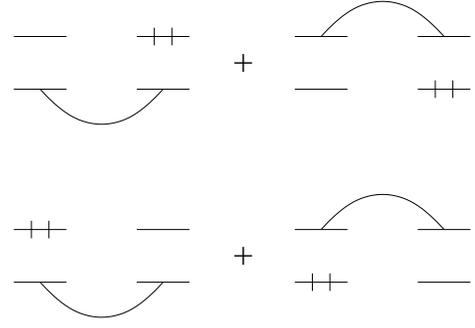}
 \caption{Two electron - two hole excitations of two weakly interacting ethylene
molecules that are reached by optical absorption from the CT exciton of Fig.~1(c).
The spin singlet diagrams shown are excited polaron-pair states corresponding 
to Eqs. (9) and (10), respectively, with each molecule having unit positive or negative charge.
}
\end{figure}

It is easy to see why charge-transfer absorption to this state from the CT exciton is 
allowed for
nonzero $t_{\perp}$. Within the exciton representation such an absorption originates from
interunit charge-transfer from either frontier MO of one unit to either frontier MO of the
second unit. \cite{Chandross99} One can then have low energy charge-transfer, from
the occupied antibonding MOs of
the intrachain excitons in Fig.~1(a) to the unoccupied antibonding MOs of the neighboring
unit, creating the low energy polaron-pair state discussed in section IV.A. This is the
P$_1$ absorption. In addition,
it is possible to have transition from the the singly occupied (doubly occupied) bonding MO of 
one unit of the forbidden exciton to the 
unoccupied (singly occupied) antibonding MO of the neighboring unit.
This second transition is clearly at higher energy, and gives the excited polaron-pair
states of  Figs. 3(a) and (b). In spite of the basic similarity between ground state
and excited state charge-transfer, there then does exist one fundamental difference between them,
viz., multiple charge-transfer absorptions will occur in the latter case, as opposed to a 
single absorption in the former.

\subsection{Finite oligomers versus polymers}

One possible criticism of the work presented above might be that the calculations are for finite 
oligomers and therefore they may not apply to infinite chains. Different research groups have
shown, for instance, that within the dipole-dipole interaction model the splitting between the
optical and the dark exciton vanishes in the infinite chain length. \cite{Clark07,Cornil98,Barford07}
We point out the following in this
context. First, the dipole-dipole interaction model is valid only when the polaron-pair configuration
is ignored ($t_{\perp} \to 0$ limit). For reasonable $t_{\perp}$ and $V_{ij}^{\perp}$, the stabilization
of the dark exciton comes predominantly from CI with the polaron-pair (see Appendix). 
Second, real PCPs are not true
infinite chains and usually consist of a distribution of conjugation lengths that are close to what we have
considered here. \cite{Basko02,Chandross97} As seen in the Appendix, the energy gap between the optical and the CT exciton does
indeed decrease with size, but there is a broad region over which this gap is nearly the same.
Third, and most important in the present context, it is not the gap between
the optical exciton and the CT exciton, but rather, {\it the gap between the polaron-pair and the CT exciton,}
that is relevant for our theory. We show in the Appendix that this second gap, corresponding to the
P$_1$ transition energy, increases weakly with increasing size.

\section{Discussions and Comparison to Experiments}

Our work provides the insight necessary to understand the various mutually contradictory 
experimental results. There occur in PCPs a series of interchain CT excitons and polaron-pairs
in the energy space below the continuum band. The lowest CT exciton occurs below the optical
exciton, and its wavefunction is a superposition of (a) the wavefunction of the lowest 
state in the exciton band of a H-aggregate, and (b) the lowest polaron-pair state 
of odd parity. 
The disorder-induced 0-0 emission, as well as the 0-1 emission
from the lowest H-aggregate state and from the CT exciton are
therefore very likely similar, since the polaron-pair component of the CT exciton 
has no dipole coupling with the ground state and should not interfere in the emission process.
The success of the H-aggregate model in explaining the PL of rrP3HT \cite{Clark07,Spano05} therefore
does not contradict the CT exciton scenario. As emphasized by others, \cite{Martini04} probing
at a variety of wavelengths is essential for understanding the complete role of morphology.  

Whether or not significant CT occurs in real systems depends on the
magnitude of $t_{\perp}$ and the relative energies of the dark exciton and the polaron-pair in the
absence of $t_{\perp}$. The demonstration of delocalized two-dimensional polarons in rrP3HT
\cite{Osterbacka00} proves that $t_{\perp}$ should be large enough for interchain electron
hopping. Similar large $t_{\perp}$ has been assumed in calculations for CN-PPV and MEH-PPV 
\cite{Wu97,Conwell98,Meng98}. As mentioned already, delayed PL in films is cited as evidence for
some polaron-pairs occurring even below the optical exciton \cite{Rothberg06,Arkhipov04} 
(it is not being implied that these polaron-pairs are generated in photoexcitation.) This would 
suggest that even though the bulk of the polaron-pairs are above the exciton, they are proximate in
energy (see also Appendix). Taken together, moderate $t_{\perp}$ and relatively low energy  
polaron-pairs indicates significant charge-transfer.

Within our theory, PA in films is from both the CT exciton and the optical exciton at the earliest
times, and predominantly from the CT exciton following this. 
The similarity between the two-chain PA in Fig.~2 and the low energy part of the experimental PA spectra of
by Sheng {\it  et al.} \cite{Sheng07} is striking. PA$_1$ in solutions is the absorption from the single-chain
1B$_u$ to the mA$_g$. As in Sheng {\it et al's} experiment for solutions, the P$_1$ absorption is missing 
in our single chain calculation. The P$_1$ absorption in films, however, is not from free polarons, but
is a charge-transfer absorption from the lowest CT exciton to the lowest polaron-pair. PA in the 1 eV
range in solutions and films appear to be identical but have slightly different origins: in films
this is the PA$_1^{\prime}$ absorption from the lowest CT exciton to a higher energy CT exciton in the
mA$_g$ - manifold. The branching of photoexcitations, as discussed by Sheng {\it  et al.}, is real,
and the instantaneous generation of P$_1$ is likely a consequence of the CT exciton being weakly
allowed in absorption due to disorder. Our theory is a straightforward extension of 
Mulliken's theory of ground state charge-transfer in a donor-acceptor
complex \cite{Mulliken52} to the case of photoinduced charge-transfer in PCP films. 
As in Mullken's theory, there
occur from the CT exciton charge-transfer absorptions (viz., P$_1$ in Fig.~2)
absent in the pure ``molecular'' components,
in addition to the weakly perturbed ``molecular absorption'' PA$_1^{\prime}$. 

Our interpretation of P$_1$ explains the absence of room temperature IRAV in Sheng {\it et al's}
photoexcitation experiment, since free charges are not generated. The weak low temperature IRAV 
may owe its origin to the polaron-pair
contributions to the CT exciton wavefunction. In the disordered case, the 
contributions by $|exc1\rangle$ and $|exc2\rangle$, and by
$|P^+_1P^-_2\rangle$ and $|P^-_1P^+_2\rangle$, respectively, to the CT exciton are different, and this
asymmetry may make weak IRAV possible. This is currently being investigated. The apparent contradiction
between ultrafast spectroscopy on the one hand, and microwave \cite{Dicker04} and 
THz spectroscopy \cite{Hendry05} 
on the other, is also understandable once it is recognized that P$_1$ is not associated with polarons.

The PA$_2$ seen in solutions is a second higher energy ``molecular absorption'', and from the above
extension of Mulliken theory, we expect a weakly perturbed PA$_2^{\prime}$ absorption in films.
We have not tried to directly evaluate this PA, as
even in single chains the understanding of PA$_2$ requires highly
sophisticated many-body calculations.
\cite{Shukla03} Similar calculations
are currently beyond our reach for the two-chain case, but the results of Tables I and II indicate
that the interchain species that is the final state of PA$_2^{\prime}$ must exist.
More interesting is the higher energy PA
peculiar to films and absent in solutions.
\cite{Yan94,Hsu94,Sheng07,Singh08} 
We have not calculated this higher energy PA for PPV oligomers, but have determined that such
an absorption from the CT exciton is found in FCI calculations on coupled ethylene and butadiene
chains. The analogy between coupled ethylenes and long PCP chains pointed out in section IV
suggest that similar high energy two electron - two hole polaron-pair state will exist also
in arbitrary PCPs. Based on the very slow polarization memory decay kinetics, the high energy PA associated 
with films has recently been ascribed to absorption from the CT exciton, \cite{Singh08} in agreement
with our prediction.

Our calculations allow us to make predictions for polarizations of the PAs. 
We predict that PA$_1^{\prime}$ in films will be polarized along the PCP chains, and that  
P$_1$ will be polarized transverse to the chains. Preliminary polarization memory measurements
are in agreement with these predictions, but more careful measurements have to be performed to
confirm that the PAs are from the same species. \cite{Vardeny08}
Finally, CT excitons have also been claimed in recent experiments on
dendritic oligothiophenes \cite{Zhang07} and in pentacene films. \cite{Marciniak07}

While the present theoretical work has focused entirely on single-component PCPs, several
experimental groups have recently discussed charge-transfer complexes (CTCs) created upon photoexcitations
of heterstructures composed of donors and acceptors. 
\cite{Morteani04,Sreearunothai06,Hwang08,Drori08}
The donor-acceptor polaron-pair as well as the CTC here are expected and found below the optical
gaps of the donor as well as the acceptor. Theoretical work on excited state absorptions from the
CTCs \cite{Drori08} is of interest and is currently being pursued.

\begin{acknowledgments}
S. M. thanks Z. V. Vardeny for suggesting this work and for the hospitality extended by
the University of Utah where this work was conceived. We are grateful to L. J. Rothberg
and C.-X. Sheng for many stimulating discussions and for sending preprints.
This work was supported by NSF-DMR-0705163.
\end{acknowledgments}

\appendix*
\section{}

In order to understand finite size effects associated with our results we have calculated the energy 
spectra near the optical gap edge for pairs of linear polyenes as well as for PPV oligomers for many different
chain lengths.
In Fig.~4(a) we show our results for pairs of 
cofacial linear polyenes for $\kappa_{\perp}=2.5$. The number of carbon atoms per chain N ranges from 10
to 70. The energy orderings are the same as in Fig.~1. We show plots of (i) the energy difference between
the optical exciton and the dark exciton, $\Delta E_{e-e}$ and (ii) the energy difference between the polaron-pair and
the CT exciton, $\Delta E_{pp-e}$.
We have chosen larger $\kappa_{\perp}$ than in Tables
I and II (weaker interchain electron-hole attraction) since for $\kappa<2$ the polaron-pair and the 
CT exciton are both too low in energy at small N, and the ionicity of the CT exciton is much larger than
in Tables I and II. For $\kappa_{\perp}=2.5$ the ionicities are comparable. Our results for $\Delta E_{e-e}$
should be compared against those obtained using the supermolecular
approach in reference \onlinecite{Beljonne00} (see Fig.~2(a) of this reference which shows results for
interchain separation of 0.45 nm). 
For the same N values, the $\Delta E_{e-e}$ are comparable. We have plotted our energy differences against
N rather than 1/N to point out that although $\Delta E_{e-e}$ indeed decreases with N, there is a broad
range of N where the decrease is slow. For real polyacetylene films we exect $\Delta E_{e-e} \neq 0$.

\begin{figure}
\centering
\vspace{0.2in}
\includegraphics[clip,width=3.375in]{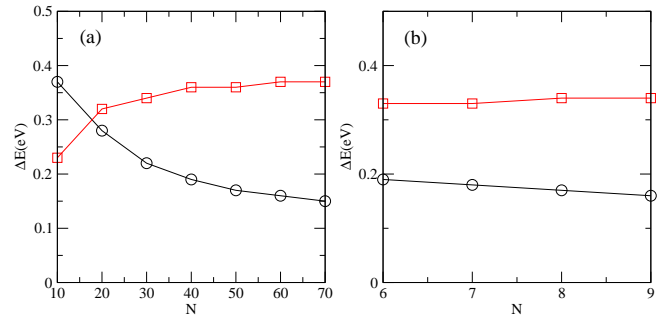}
\caption{Energy difference between the optical and CT exciton (circles), and between the 
polaron-pair and the
CT exciton (squares) in pairs of (a) linear polyenes, (b) PPV oligomers.
}
\end{figure}

The plots for $\Delta E_{e-e}$ and $\Delta E_{pp-e}$ are not completely independent. As seen in the figure,
decreasing $\Delta E_{e-e}$ is accompanied by increasing $\Delta E_{pp-e}$, which is a signature
that the bulk of the stabilization of the CT exciton is coming from CI with the polaron-pair
(the CI decreases with increasing energy of the of the polaron-pair). Again, $\Delta E_{pp-e}$ is
nearly the same over a broad range of N.

In Fig.~4(b) we have shown the same results for PPV oligomers with $\kappa_{\perp}=2$ , 
with N now the number of units as opposed to number of carbon atoms. Again, our results for $\Delta E_{e-e}$
should be compared against Fig.~3 of reference \onlinecite{Cornil98}, where, however, the calculations
go up to 7 units only. As in Fig.~4(a), the plots against N (as opposed against 1/N) 
make the slow variation of the energy differences against size clear. Decreasing $\Delta E_{e-e}$ 
is accompanied by increasing $\Delta E_{pp-e}$,
as in Fig.~4(a).
 
Finally, it is not being implied that the actual magnitudes of the calculated energy differences should
be taken seriously. The quantitative aspects of the calculations depend to a large extent on the
parametrization of $V_{ij}$ and $V_{ij}^{\perp}$, which are not known. Equally importantly, the
effects of background polarization are difficult to estimate. It may, however, be significant
that our parametrization \cite{Chandross97} of $V_{ij}$ has given the most accurate estimations
of exciton energies and exciton binding energies in a different family of $\pi$-conjugated
systems, viz., single-walled carbon nanotubes, to date. \cite{Wang06}

\end{document}